\documentclass{iaus}
\usepackage{graphicx}

\title[Structural Parameters of Bulges, Bars and Discs] %% give here short title %%
{The Structural Parameters of Bulges, Bars and Discs in the Local Universe}

\author[Dimitri Gadotti]   %% give here short author list %%
{Dimitri Alexei Gadotti}

\affiliation{Max-Planck-Institut f\"ur Astrophysik \break Karl-Schwarzschild-Str. 1, D-85748
Garching bei M\"unchen, Germany \break email: dimitri@mpa-garching.mpg.de}

\pubyear{2007}
\volume{245}  %% insert here IAU Symposium No.
\pagerange{??--??}
\date{?? and in revised form ??}
\jname{Formation and Evolution of Galaxy Bulges}
\editors{M. Bureau, E. Athanassoula \& B. Barbuy, eds.}
\begin{document}

\maketitle

\begin{abstract}
Image decomposition of galaxies is now routinely used to estimate the structural parameters of galactic
components. In this work, I address questions on the reliability of this technique. In particular, do bars
and AGN need to be taken into account to obtain the structural parameters of bulges
and discs? And to what extent can we trust image decomposition when the physical spatial resolution is
relatively poor? With this aim, I performed multi-component (bar/bulge/disc/AGN) image decomposition of a
sample of very nearby galaxies and their artificially redshifted images, and verified the effects of
removing the bar and AGN components from the models. Neglecting bars can
result in a overestimation of the bulge-to-total luminosity ratio of a factor of two, even if the resolution
is low. Similar effects result when bright AGN are not considered in the models, but only when the
resolution is high. I also show that the structural parameters of more distant galaxies can in general be
reliably retrieved, at least up to the point where the physical spatial resolution is $\approx$ 1.5 Kpc, but
bulge parameters are prone to errors if its effective radius is small compared to the seeing radius, and might suffer
from systematic effects. I briefly discuss the consequences of these results to our knowledge of the stellar mass
budget in the local universe, and finish by showing preliminary results from a large SDSS sample on the dichotomy
between classical and pseudo-bulges.
\keywords{galaxies: bulges, galaxies: evolution, galaxies: formation, galaxies: fundamental parameters,
galaxies: photometry, galaxies: structure}
%% add here a maximum of 10 keywords, to be taken form the file <Keywords.txt>
\end{abstract}

\firstsection % if your document starts with a section,
              % remove some space above using this command.
\section{Introduction and Motivation}

Parametric decomposition of galaxy images has become a popular tool to estimate the structural parameters
of different galactic components, particularly bulges and discs.
Through this sort of analysis, one is also able to determine the relative importance of the bulge component,
with parameters such as the bulge-to-total luminosity ratio B/T, one of the major features that
define the \cite{hub26} sequence. It thus provides indispensable means to investigate the formation
and evolution of galaxies, and the origin of the Hubble sequence, some of the key subjects in current
astrophysical research. Recently, the focus of studies using image decomposition shifted
from relatively small samples of very nearby galaxies, where the fits can be done on a more careful,
individual basis (e.g., \cite[de Jong 1995]{dej95}), to include large samples of more distant galaxies,
using automated procedures, that allow solid statistical analyses (e.g., \cite[Allen \etal\ 2006]{all06}).
So far, most studies have ignored other components, such as bars (but see
\cite[Laurikainen \etal\ 2004, 2005]{lau04,lau05}), although the majority of disc galaxies host bars
that usually contain a significant fraction of the galaxy total stellar mass. It is unclear how the
parameters of bulge and disc are affected when bars are ignored. A similar issue concerns galaxies
that host bright, type 1 AGN: if the light contribution from the AGN is not modelled, the structural
parameters obtained might be incorrect. It is also not clear if the low physical spatial resolution
usually found in images of more distant galaxies introduces any bias in the parameters recovered.
To shed light on these issues, I performed multi-component image decomposition of a
sample of very nearby galaxies and their artificially redshifted images, and verified the effects of
removing the bar and AGN components from the models. The results of these tests are shown in the next
section [the reader is referred to \cite{gad07} for full details on this work].

\section{Tests on Image Decomposition}

The images used here are from \cite{gad06}. These are V and R images of 17 galaxies
at $z\approx0.005$, most of them hosting bars and AGN. Image decomposition is done using {\sc budda}
v2.1 (see \cite[de Souza, Gadotti \& dos Anjos 2004]{des04}). All components are modelled as sets of
concentric ellipses, with constant ellipticity and position angle. Bulges are described with a S\'ersic
luminosity profile and discs are exponential. Bars are described with boxy ellipses, also with
a S\'ersic profile, and the AGN contribution is modelled using a circular Moffat profile.
When one takes advantage of the full capabilities of the code (i.e., all necessary components)
on these images, the results show a fairly good agreement with other studies. In addition, several
known scaling relations are reproduced.

\begin{figure}
\center
\includegraphics[width=4cm,clip=true]{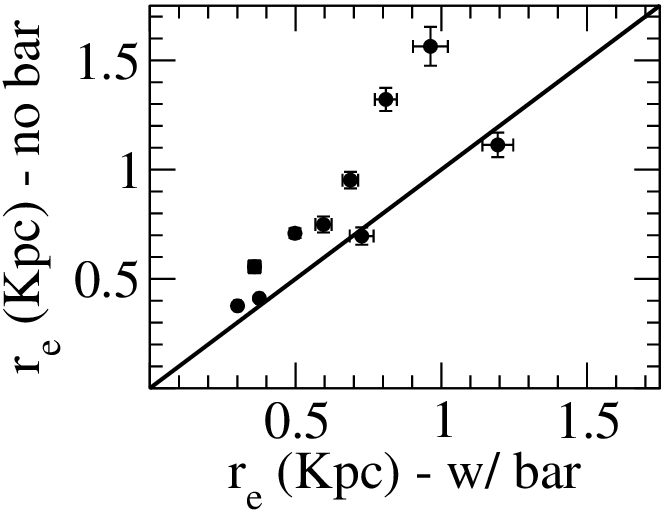}\hskip0.3cm
\includegraphics[width=9cm,clip=true]{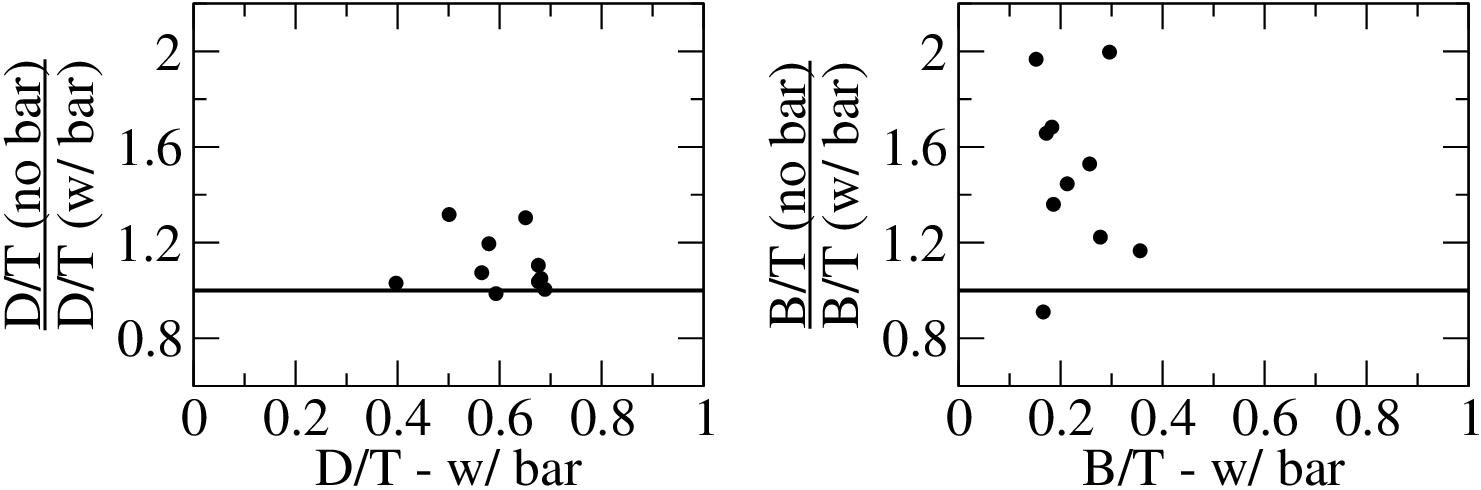}
\caption{Left: effective radius of the bulge, as estimated when bars are not included in the models,
plotted against the same parameter when bars are taken into account. The two remaining panels show the
relative overestimation of the disc-to-total and bulge-to-total luminosity fractions when bars are
neglected, plotted against the corresponding parameters when models include bars. The solid lines
depict a perfect correspondence. This figure shows that when bars are ignored bulges get bigger, in a
way to accommodate the light from the bar. Disc models are also affected, but the effect is stronger
for bulges: B/T can be overestimated by a factor of two.}\label{nobar}
\end{figure}

When one does not include the bar component in the models, the major effect is seen in the bulge models
recovered: bulges get bigger, in a way to accommodate light from the bar. As a result, B/T is
overestimated, on average, by 50\%, but this overestimation can reach a factor of 2. Disc models
are also systematically affected, but to a lesser extent: D/T is overestimated, on average, by 10\%, with a
maximum overestimation of 30\% (see Fig. \ref{nobar}). Similar effects
(albeit restricted to the bulge component) result when bright, type 1 AGN
are not considered in the models: B/T and the S\'ersic index of the bulge $n$ can be overestimated by
factors of 2 and 4, respectively.

\begin{figure}
\center
\includegraphics[width=13.3cm,clip=true]{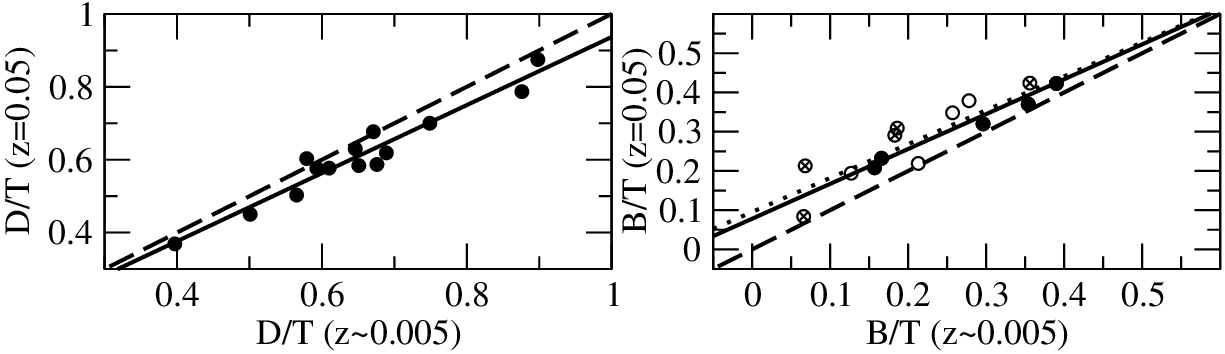}
\caption{Disc and bulge luminosity fractions, as determined with the redshifted images, plotted
against the same parameters obtained with the original images. The dashed lines
indicate a perfect correspondence. For B/T, filled circles correspond to those galaxies
where the effective radius of the bulge in the redshifted image is larger than the seeing radius, the empty circles
correspond to those galaxies where it is similar to the seeing, and the circles with crosses correspond to those galaxies
where it is smaller than the seeing. The solid and dotted lines are linear fits to the data. For B/T, the dotted
line is a fit to all data points, while the solid line corresponds to a fit where the circles with crosses were excluded.
One sees that, in general, structural parameters can be reliably retrieved through
image fitting even in a low resolution regime. Nevertheless, bulge parameters are prone to errors if its
effective radius is small compared to the seeing radius, and might suffer from systematic effects.}\label{hz}
\end{figure}

To verify if these results hold when the physical spatial resolution of the images is relatively low,
I degraded the original images, thus artificially redshifting them to $z=0.05$. The resolution in
these images is 1.5 Kpc, which is the typical resolution of SDSS at this redshift.
Such a resolution is also typical in other works as well, and is what can be achieved at $z\approx1$
with HST. As with the original images, the redshifted images were first used to perform a complete
decomposition, and then to test the effects of removing the bar and AGN components from the models.
Figure \ref{hz} shows that, in general, structural parameters can be reliably retrieved through
image fitting even in a low resolution regime. This is particularly true for the disc parameters.
However, bulge parameters are prone to errors if its effective radius $r_e$ is
small compared to the seeing radius, and B/T might suffer from systematic effects. B/T seems to be
systematically overestimated, on average, by 0.05, i.e., 5\% of the total galaxy luminosity, even when
$r_e$ is similar to the seeing radius.

When bars are excluded from the models to fit the redshifted images, one still sees similar effects as
with the original images, namely, a significant overestimation of B/T and $r_e$,
if bars are prominent. The change in $r_e$ is even
more pronounced in the redshifted images, likely because the geometrical properties of the bulge are substantially
smoothed by the seeing, making it more difficult to constrain its properties. For less prominent bars these
effects are reduced in strength. The effects caused by not taking into account bright
AGN, i.e., the overestimation of B/T and $n$, do {\em not} occur with the redshifted images,
as the AGN contribution is smeared out by the seeing.

\section{The Stellar Mass Budget in the Local Universe}

When bars are not taken into account, the amount of mass
in stars in bulges and discs is overestimated, and the excess is an indication of the amount of mass in stars
that reside in bars. \cite{dri07} estimated the $z\approx0$
stellar mass content in classical bulges and discs through image decomposition of $\approx10^4$ galaxies,
with a spatial resolution similar to that of the redshifted images above.
Bars are not taken into account in the fitted models, but these authors made a thorough quality control,
removing poor fits. However, it is usually the case that, even when there is no bar in the model,
when fitting a barred galaxy, one gets an acceptable (though wrong) fit, essentially because the
bulge model is distorted. Hence, with reasonable assumptions,
one can use their results to obtain a {\em rough} estimate of what can be the stellar
content in bars, assuming that the biases produced in a low resolution regime and by ignoring bars, as found above,
can be used in this case to obtain the true bulge and disc luminosity fractions.
Thus, the stellar content in classical bulges and discs is found to be
$\approx13.5\%$ and $\approx58.5\%$, respectively. And the stellar content in bars is $\approx12\%$.
Although these are rough estimates, that need to be confirmed by further studies, they open up the possibility of
bars being as important as bulges in what concerns their stellar mass content in the local universe.

\section{The Classical vs. Pseudo Bulge Dichotomy}

\begin{figure}
\center
\includegraphics[width=9cm,clip=true]{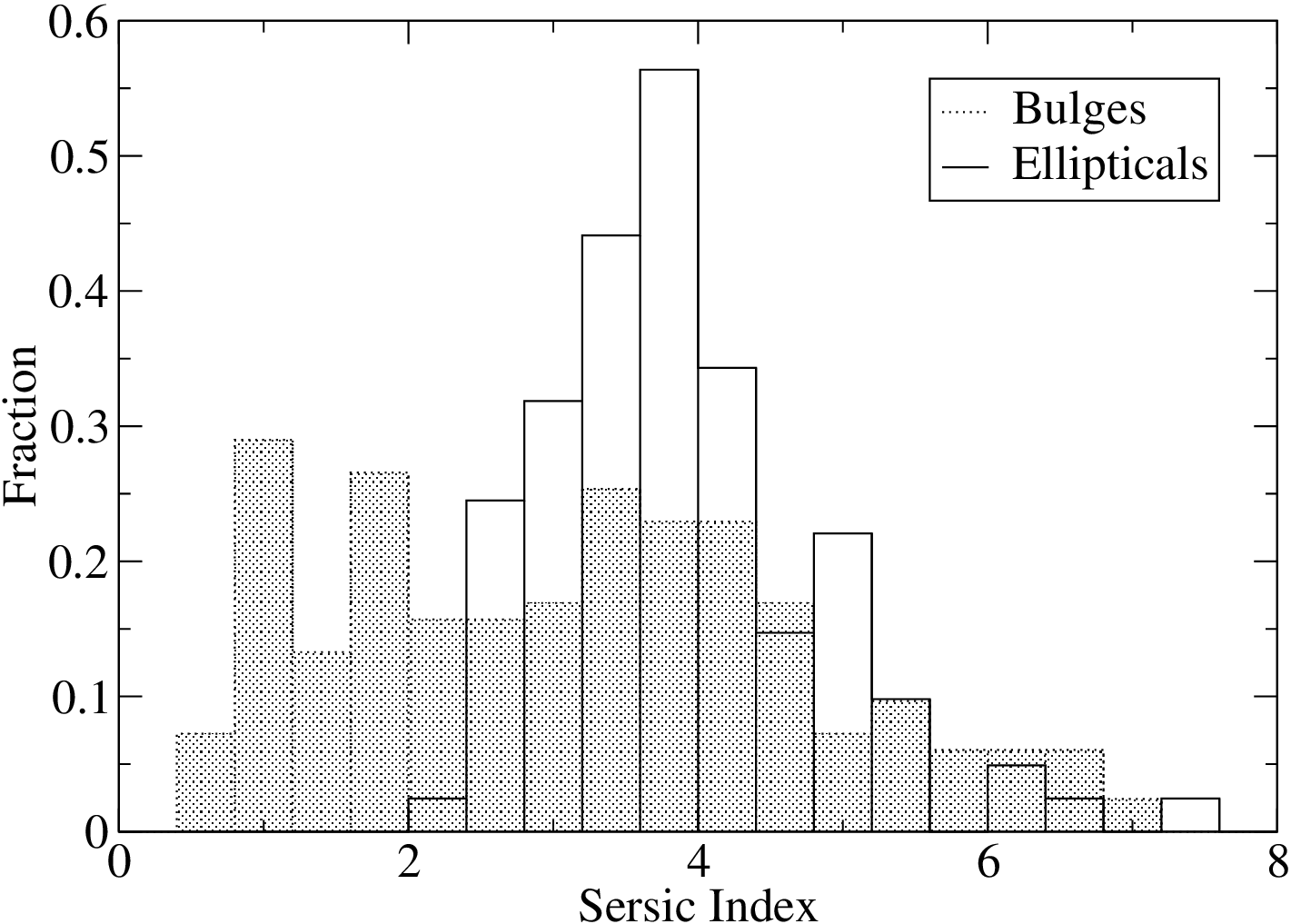}
\caption{The distribution of the S\'ersic index of bulges and ellipticals in a sample of about 300
galaxies (of which $\approx200$ are disc galaxies) from the SDSS.
Bulges show a remarkably different distribution as compared to ellipticals, with the family of pseudo-bulges
(i.e., those with S\'ersic index less than about 2) standing out clearly.}\label{ndistr}
\end{figure}

Figure \ref{ndistr} shows the distribution of the S\'ersic index of bulges and ellipticals in a sample of
about 200 disc galaxies and 100 ellipticals from the SDSS, obtained using {\sc budda} v2.1
(see \cite[Gadotti \& Kauffmann 2007]{gadkau}). One clearly sees
the dichotomy between classical bulges, with $n\approx4$, and pseudo-bulges, with $n\lesssim2$. This figure
indicates that the fraction of pseudo-bulges at $z\approx0$ is $\approx35\%$.

\begin{acknowledgments}
I would like to thank Martin Bureau and the organisers for this very fruitful meeting.
It is a pleasure to thank Guinevere Kauffmann for her active role in this work, with insightful comments
and suggestions, and Ronaldo de Souza, for his help in the making of the new {\sc budda} version.
DAG is supported by the Deutsche Forschungsgemeinschaft priority program 1177 (``Witnesses of Cosmic
History: Formation and evolution of galaxies, black holes and their environment''), and the Max Planck
Society.
\end{acknowledgments}

\end{document}